\documentclass[aps,prb,reprint,groupedaddress,showpacs,amsmath,amssymb]{revtex4-1}
\usepackage{graphicx}
\usepackage{dcolumn}
\usepackage{bm}
\usepackage{float}
\usepackage{url}
\usepackage{amssymb}
\usepackage{latexsym}
\usepackage{epsf}
\usepackage{notoccite}
\usepackage{euscript}
\usepackage{xcolor}

\begin{document}
\title{Temperature evolution of polaron dynamics and Jahn-Teller distortion modes in strongly correlated La$_{0.67}$Ca$_{0.33}$MnO$_{3}$ manganite film}

\author{Naween Anand}
\altaffiliation{National High Magnetic Field Lab, FSU, Tallahassee , FL 32310-3706, USA}
\affiliation{Department of Physics, University of Florida, Gainesville, FL 32611-8440, USA}
\author{Naveen Margankunte}
\altaffiliation{Senior Test Engineer, Areva Solar, 2585 E Bayshore Rd, Palo Alto, CA 94303, USA}
\affiliation{Department of Physics, University of Florida, Gainesville, FL 32611-8440, USA}
\author{Hyoungjeen Jeen}
\altaffiliation{Department of Physics, Pusan National University, Busan, South Korea}
\affiliation{Department of Physics, University of Florida, Gainesville, FL 32611-8440, USA}
\author{A. F. Hebard}
\affiliation{Department of Physics, University of Florida, Gainesville, FL 32611-8440, USA}
\author{Amlan Biswas}
\affiliation{Department of Physics, University of Florida, Gainesville, FL 32611-8440, USA}
\author{D. B. Tanner}
\affiliation{Department of Physics, University of Florida, Gainesville, FL 32611-8440, USA}
\date{\today}

\begin{abstract}
Reflectivity as a function of temperature for the La$_{0.67}$Ca$_{0.33}$MnO$_{3}$ (LCMO) film has been measured across the metal-insulator phase transition. The optical properties and their temperature dependence were determined in the infrared and visible range by fits to a Drude-Lorentz model, using exact formula for the thin film optics and the measured properties of the substrate. The phonon modes were identified and verified with lattice dynamical calculations for the ideal and distorted perovskite structure of the material. The optical conductivity shows agreement with the double exchange mechanism in conjunction with the Jahn-Teller distortion term in the Hamiltonian. Low temperature metallic phase is dominated by large polaron dynamics, a key component of electron-orbital coupling in a strongly corrrelated system. Free carrier dynamics in the metallic phase is described in terms of coherent heavy polaronic motion in the DC limit with incoherenet and asymmetric polaronic background in the mid-IR range. The strength and line width of Jahn-Teller modes has been discussed across the phase transition and their temperature evolution is qualitatively discussed on account of existing electron-phonon coupling. The localized Holstein polaron formation in the high temperature insulative phase is identified as optical conductivity peaks in the visible range above the critical temperature.
\end{abstract}
\maketitle

\section{INTRODUCTION}
The field of condensed matter physics is mainly driven by the ever increasing need for new exotic materials and phases whose properties could be exploited to bring about further technological advancement. Strongly correlated 3d transition metal oxides (TMO) have drawn enormous attention among scientific communities in last few decades. Due to the localized nature of the 3d orbitals, these transition metal oxides are known to possess strong electron-electron correlations making them quite an interesting but challenging field of research in condensed matter physics. The doped rare earth manganite system represents one such class of materials exhibiting rich ground states which results from the interplay between spin, charge, lattice and orbital degrees of freedom.\cite{Zener,Millismueller,Tokura,Schiffer,Millisnature} Phase diagrams of numerous manganite families indicate several competing phases at the domain boundaries within the crystal which leads to the formation of intrinsically inhomogeneous systems.\cite{Dagotto,dagottonanoscale,Dagottoscience} These manganites have been a potential candidates for non-volatile Magnetoresistive Random Access Memory (MRAM) modules due to an increased areal density. Thin films or single crystals of these manganites could also be used in making magnetic sensors, bolometric detectors and as a chemical catalyst in automobile industry. Besides, their ferroelectric properties also make them conducive for non-volatile ferroelectric field effect device application.\cite{coltokura,Ramirez,Coey}

\begin{figure}[H]
\centering
\includegraphics[width=3.3 in,height=3.5 in,keepaspectratio]{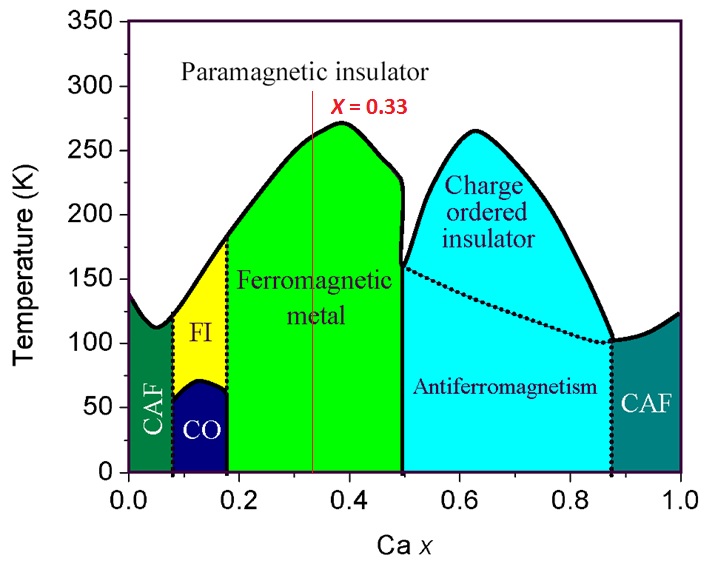}

\caption{\label{phasediagram}(Color online) Phase diagram of La$_{1-x}$Ca$_{x}$MnO$_{3}$ doped manganite as a function of doping parameter $x$.}
\end{figure}

The general chemical formula for the series is RE$_{1-x}$A$_{x}$MnO$_{3}$ with RE$^{3+}$ a rare earth trivalent cation and A$^{2+}$ an alkaline divalent cation. Oxygen is in O$^{2-}$ state, and the relative fraction of Mn$^{4+}$ to Mn$^{3+}$ is regulated by the composition $x$ ($\frac{Mn^{4+}}{Mn^{3+}} = \frac{x}{(1-x)}$). The undoped parent manganite compound LaMnO$_{3}$ is an insulating paramagnetic material with orthorhombic crystal structure that orders antiferromagnetically at around 140 K. A-type magnetic ordering has been reported in previous neutron diffraction studies as the magnetic moments lie in the $ac$ plane and possess ferromagnetic alignment along $a$-axis while successive planes along $b$-axis are aligned antiparallel.\cite{Huang,Wollan} Similarly for CaMnO$_{3}$, another orthorhombic insulating paramagnetic material undergoing antiferromagnetic phase transition around 130-K has G-type ordering, meaning all moments aligned in antiferromagnetic order in all three directions.\cite{Wollan}. Several hole doped manganese perovskites such as (La, Ca)MnO$_{3}$, (La, Sr)MnO$_{3}$ and (La, Ba)MnO$_{3}$ were synthesized and studied several decades ago,\cite{Jonker, Santen} displaying ferromagnetic properties along with anomalies in the conductivity at the Curie temperature. The critical temperature were strongly dependent on the extent of hole doping which tunes the relative percentage of  Mn$^{4+}$ ion over Mn$^{3+}$ ion in the system. However depending on the growth conditions, reported properties for these compounds have varied dramatically both in terms of the crystal structure, transport properties and magnetic ordering. Nonetheless, strong upsurge in the research interest only came about after the remarkable discovery of the colossal magnetoresistance (CMR) effect in 1994 by Jin et al.\cite{Jin} The very name of the phenomenon originates from the observation of the thousand fold change in the resistivity of the LCMO films near 77 K in the presence of applied magnetic field of 5 T. The material was epitaxial films of La$_{0.67}$Ca$_{0.33}$MnO$_{3}$ grown on LaAlO$_{3}$ substrates by laser ablation process. These manganites exhibit various ground states depending on the cation doping. Based on magnetization and resistivity, the inferred phase diagram of La$_{1-x}$Ca$_{x}$MnO$_{3}$, feature a number of distinct phases as shown in Fig.~\ref{phasediagram}.\cite{Schiffer,Millisnature} The important electrons are the Mn d-electrons, of which there is $(4-x)$ in these compounds per formula unit.  Changing the carrier concentration produces a variety of phases, which may be characterized by their magnetic, transport and charge-ordering properties.

At $x=0$, the material is insulating at all temperatures but below 140 K it goes to antiferromagnetic phase. In fact the ground state remains insulating for $x \leq 0.2$ but the magnetic order changes in a complicated and still controversial way. This sequence of phases is denoted by canted anti-ferromagnetic and ferromagnetic insulator. One also sees that charge order develops inside the ferromagnetic insulator phase which is a periodic pattern of Mn sites in different valence states.  Around $x = 0.2$, ground state changes from insulating to metallic. For $0.2 \leq x \leq 0.5$, ground state is ferromagnetic and then for $x \geq 0.5$, the ground state again becomes insulating and antiferromagnetic, and is in addition charge ordered.  For a broad range of doping, these materials have a paramagnetic insulator to ferromagnetic metallic transition upon cooling which is accompanied by a sharp drop in resistivity. Although the ferromagnetic transitions are in general very sensitive to applied fields, the very existence of the metal-insulator transition along with magnetic ordering is considered intriguing. This unusual correlation between magnetism and transport properties was first explained in terms of ``double exchange'' mechanism.\cite{Zener}  Subsequent studies on physical origination of CMR properties in manganites revealed the existence of the electron-phonon coupling due to the presence of dynamic Jahn-Teller distortion in manganites.\cite{Millis,Millisnature} The inclusion of JT coupling term along with DE term in the Hamiltonian provided more accurate prediction of Curie temperature and more realistic estimation of resistivity around T$_{c}$. In this article, a thin film sample of La$_{0.67}$Ca$_{0.33}$MnO$_{3}$ is studied which shows a paramagnetic insulating to a ferromagnetic metallic phase transition at T$_{M-I}\approx260$ K. The subsequent impact on the phonon structure due to the phase transition is studied through temperature dependent optical and transport measurements.

\section{EXPERIMENTAL PROCEDURES}
\subsection{Film growth process}
Pulsed laser deposition (PLD) technique is used to grow epitaxial film of La$_{0.67}$Ca$_{0.33}$MnO$_{3}$ (LCMO) on NdGaO$_{3}$ (NGO) substrate with (001) as exposed surface for deposition. High-power KrF excimer laser pulses were directed into growth chamber on a rotating polycrystaline LCMO target creating a supersonic jet of particles (plume) normal to the target surface. A clean NGO substrate was mounted on a Haynes alloy heater plate using Ag paint. The substrate temperature was set to 100$^\circ$C to allow the silver paint to harden and then the deposition chamber was evacuated to a pressure of 10$^{-6}$ mbar using a turbomolecular pump. The substrate temperature was then set to 820$^\circ$C at the ramp rate of 20$^\circ$C/min. When temperature reached around 300$^\circ$C, the turbo speed was set to 277 Hz from the maximum turbo frequency of 820 Hz. As frequency dropped to 600 Hz, oxygen was infused into the deposition chamber with the pressure maintained at 450 mTorr. When the temperature of the substrate increased to around 700$^\circ$C, the polycrystalline LCMO target was preablated for 10 min at a repetition rate of 10 Hz. The laser energy output was maintained around 480 mJ with energy fluence of about 2.0 J-cm$^{-2}$. The films were deposited at the rate of 0.05 nm/s at 820$^\circ$C. It was grown for about 45 min and the nominal thickness of the films were estimated to be around 140 nm. These growth parameters were optimized in order to obtain a metal-insulator transition temperature T$_{M-I}$ close to that observed in bulk compound with a minimum transition width at T$_{M-I}$. Such an optimization is
crucial since the properties of thin films of this compound vary markedly depending on the growth conditions. 

\subsection{Crystal structure analysis}
\begin{figure}[H]
\centering
\includegraphics[width=3 in,height=3 in]{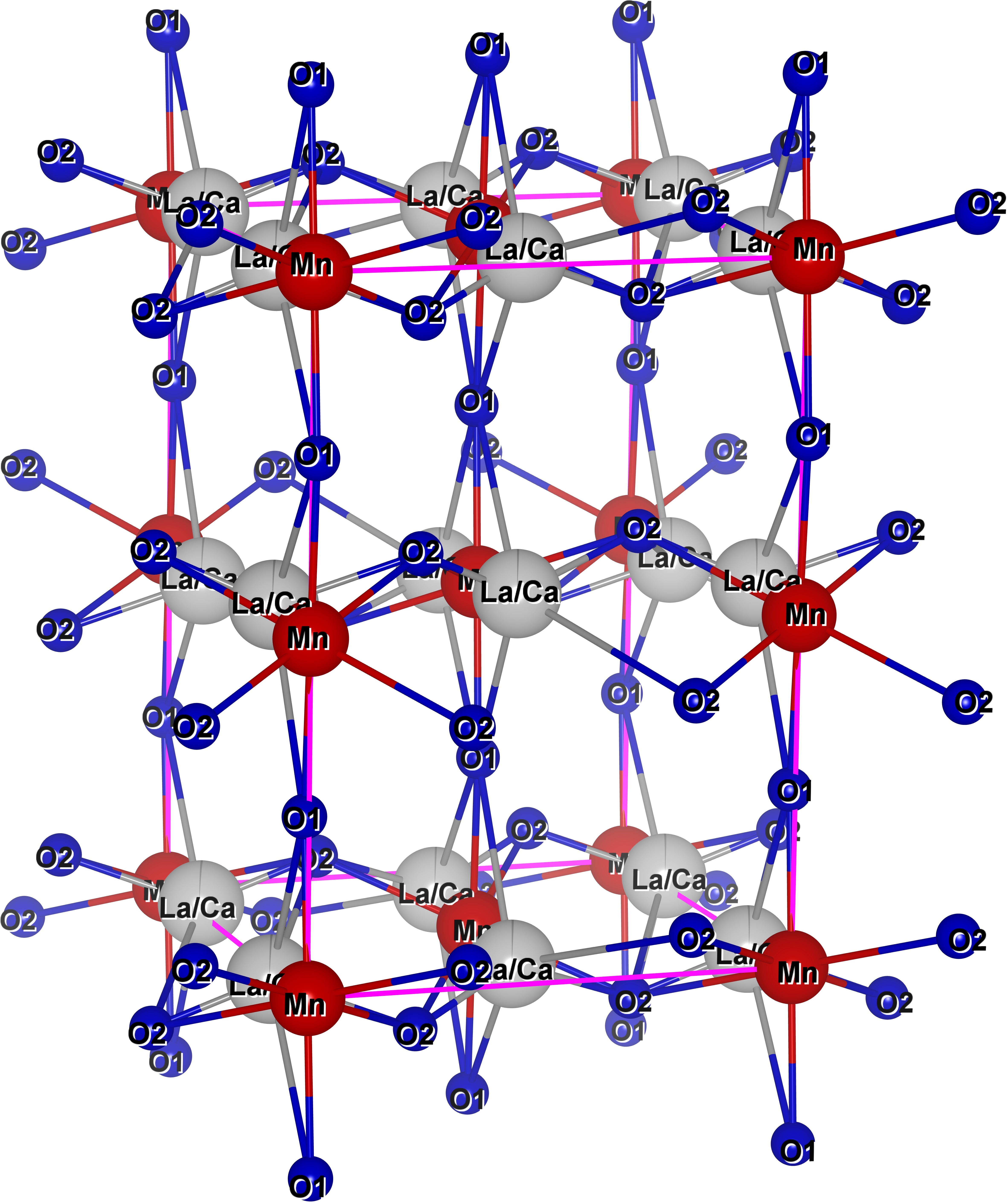}
\caption{\label{LCMO} (Color online) Orthorhombic crystal structure of La$_{0.67}$Ca$_{0.33}$MnO$_{3}$ at room temperature.}
\end{figure}

At room temperature, the film has orthorhombic unit cell with P$_{nma}$ (No. 62) space group symmetry as shown in Fig.~\ref{LCMO}\cite{Blasco,Qiyun}. Lattice parameters are as given

\begin{center}
 $a$=5.454~A$^{\circ }$; \;\;\;\;\;\;\;\;\;  $b$=5.471~A$^{\circ }$; \;\;\;\;\;\;\;\;\; $c$=7.708~A$^{\circ }$ \;\;\;\;\;\;\;\;\;  $\alpha$ =$\beta$=$\gamma$ =90$^{\circ }$
\\
Mn-O1=1.95~A$^{\circ }$; \;\;\;\; Mn-O2=1.96~A$^{\circ }$;
 \\
Mn-O1-Mn=157.9$^{\circ }$; \;\;\;\;  Mn-O2-Mn=160.0$^{\circ }$
\end{center}

\begin{figure}[H]
\centering
\includegraphics[width=2.2 in,height=2.2 in]{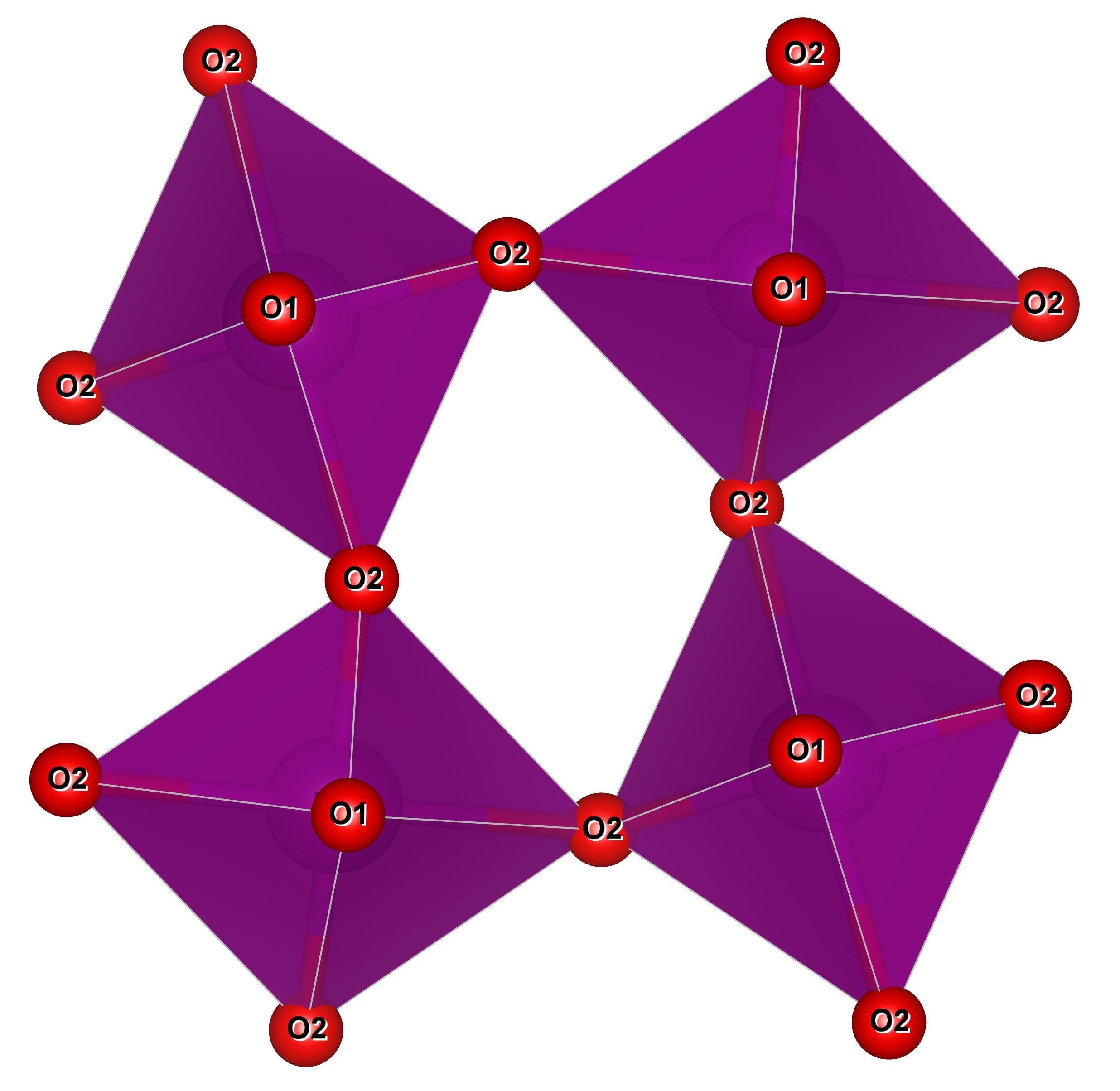}
\caption{\label{Octahedra2} (Color online) MnO$_{6}$ octahedral building blocks of orthorhombic La$_{0.67}$Ca$_{0.33}$MnO$_{3}$ at room temperature.}
\end{figure}

This structure as shown in Fig.~\ref{Octahedra2} could also be visualised in terms of distorted MnO$_{6}$ octahedral unit as building blocks which essentially dictates the transport and magnetic properties of the system. 

\vspace{-8pt}
\begin{figure}[H]
\centering
\includegraphics[width=2.8 in,height=2.8 in]{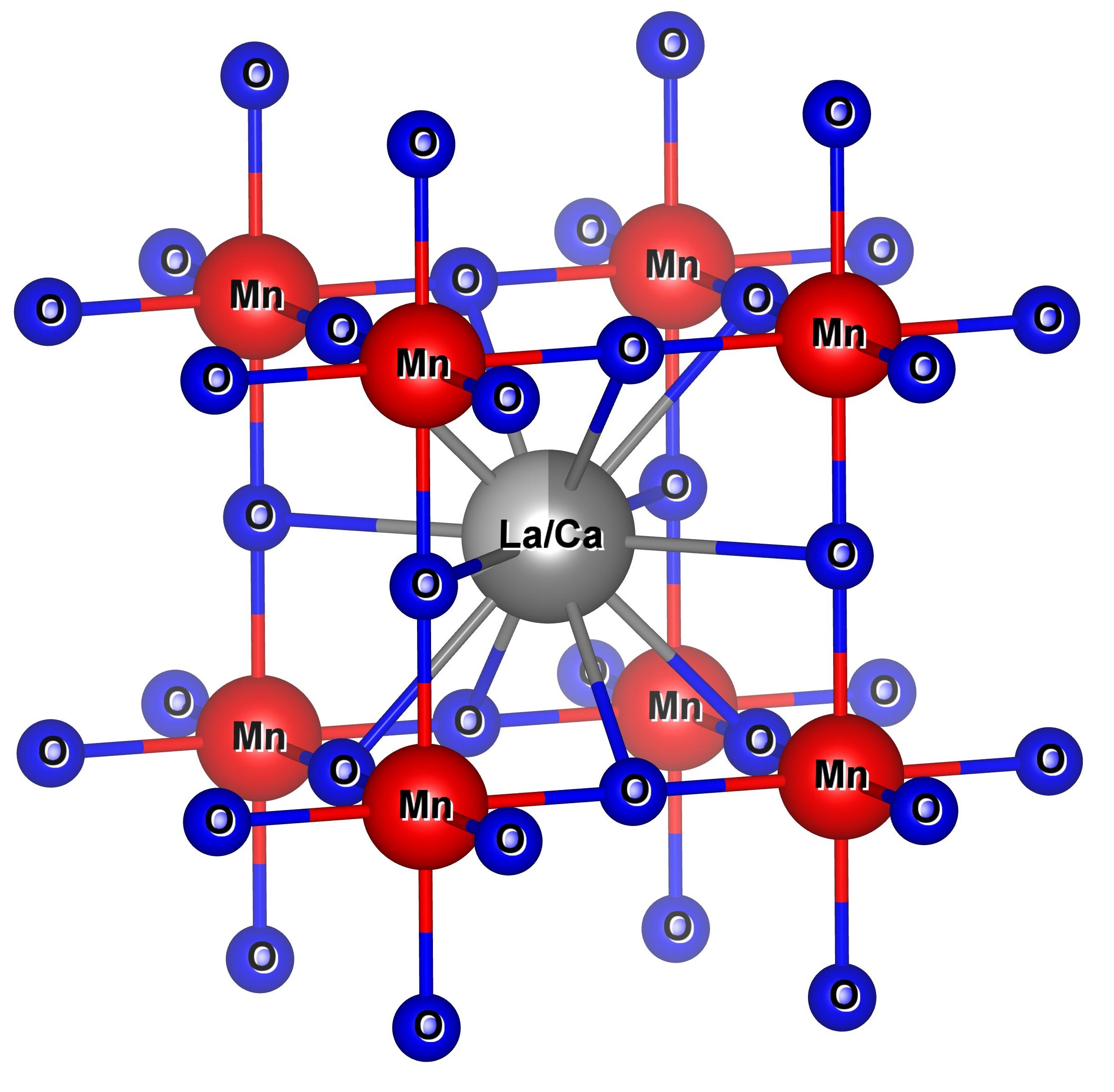}
\caption{\label{LCMOlowtemperature} (Color online) Cubic crystal structure of La$_{0.67}$Ca$_{0.33}$MnO$_{3}$ at low temperature.}
\end{figure}

In this distorted perovskite structure, one can notice two different oxygen sites labelled as O1 and O2.  In MnO$_{6}$ octahedral unit, four oxygen atoms forming the square in the horizontal plane are of one kind (O2 site) whereas one above and one below this square plane is of the second kind (O1 site). Additionally, the Mn-O bond length and Mn-O-Mn bond angle for both sites are slightly different. LCMO undergoes through a structural phase transition which is accompanied by electronic and magnetic phase transition discussed in later sections. Low temperature LCMO systems have ideal perovskite cubic structure ($\tau\approx1$)\ref{Eq1} as shown in Fig.~\ref{LCMOlowtemperature} with P$_{m\bar{3}m}$ (No. 221) space group symmetry and the structural parameters are as follows.\cite{Kolat,Atalay,JinChen}

\begin{center}
 $a$=3.873~A$^{\circ }$; \;\;\;\;\;\;\;\;\;  $b$=3.873$^{\circ }$; \;\;\;\;\;\;\;\;\; $c$=3.873~A$^{\circ }$; \;\;\;\;\;\;\;\;\;  $\alpha$ =$\beta$=$\gamma$ =90$^{\circ }$
\\
Mn-O=1.94~A$^{\circ }$; \;\;\;\; Mn-O-Mn=180$^{\circ }$
\end{center}

\begin{figure}[H]
\centering
\includegraphics[width=2.2 in,height=2.2 in]{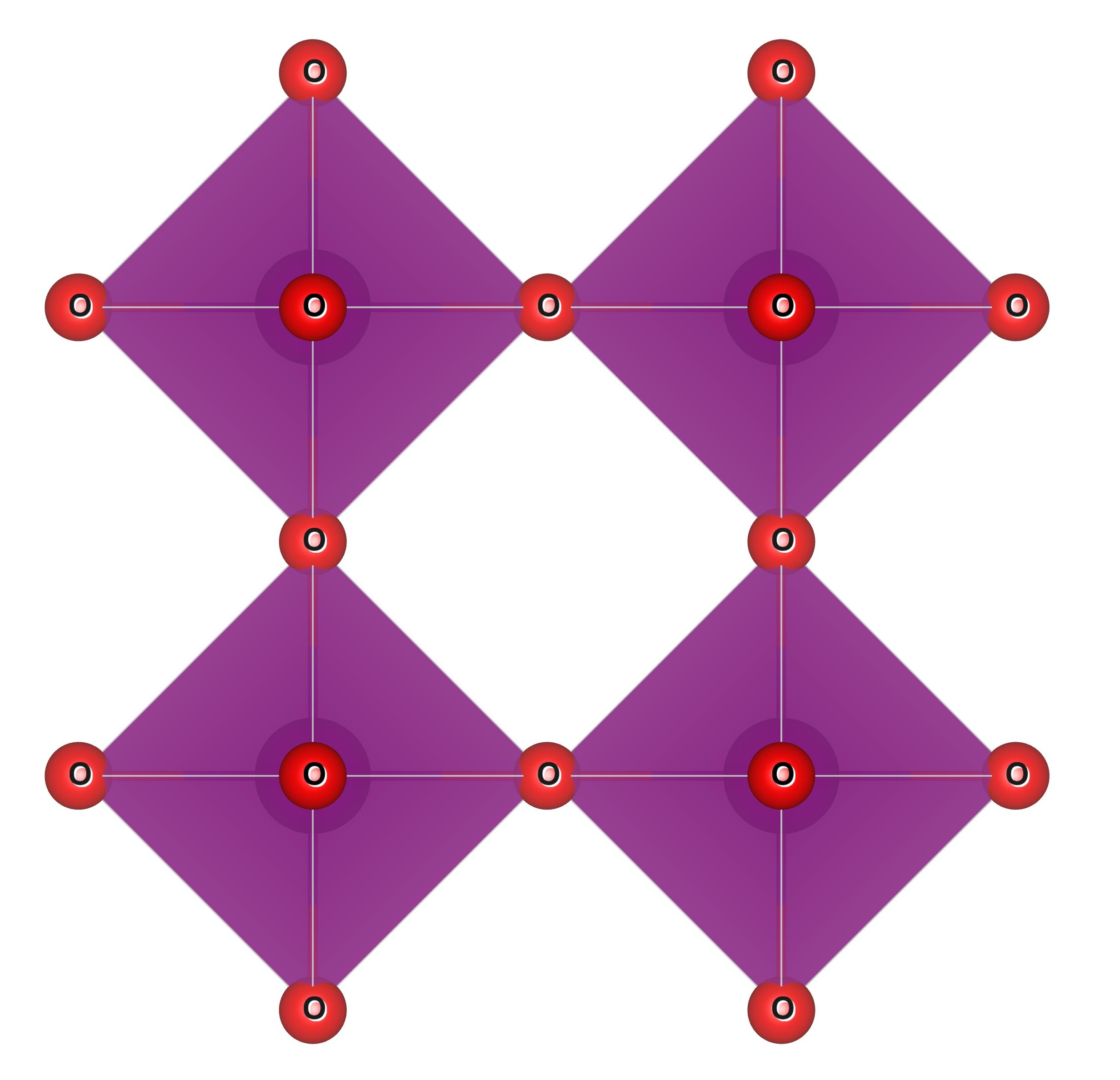}
\caption{\label{octalow} (Color online) MnO$_{6}$ octahedral building blocks of cubic La$_{0.67}$Ca$_{0.33}$MnO$_{3}$ at low temperature.}
\end{figure}

We notice that in ideal perovskite phase, all oxygen sites are identical and Mn-O bond length is slightly smaller as depicted in Fig.~\ref{octalow} showing undistorted MnO$_{6}$ octahedral chain. This greatly affects the orbital overlap between atoms and thereby the strength of spin exchange correlation between Mn ions. The extent of distortion in ABO$_{3}$ perovskite structure is estimated with the Goldsmidt tolerence factor $\tau$, given by\cite{goldschmidt}  

\begin{equation}
\label{Eq1}
\tau =\frac{\left\langle r_{A}\right\rangle+\left\langle r_{O}\right\rangle}{\sqrt{2}(\left\langle r_{B}\right\rangle+\left\langle r_{O}\right\rangle)} 
\end{equation}

where $\left\langle r_{A}\right\rangle$, $\left\langle r_{B}\right\rangle$ and $\left\langle r_{O}\right\rangle$ are mean radii of the ions A, B and oxygen ion positions respectively. The manganite system retains the perovskite structure if the tolerence factor stays in the limit of 0.75$\leq\tau\leq$1, being unity in an ideal perovskite case.\cite{Jonker,JONK,HUAN}

\begin{figure}[H]
\centering
\includegraphics[width=2.6 in,height=2.6 in]{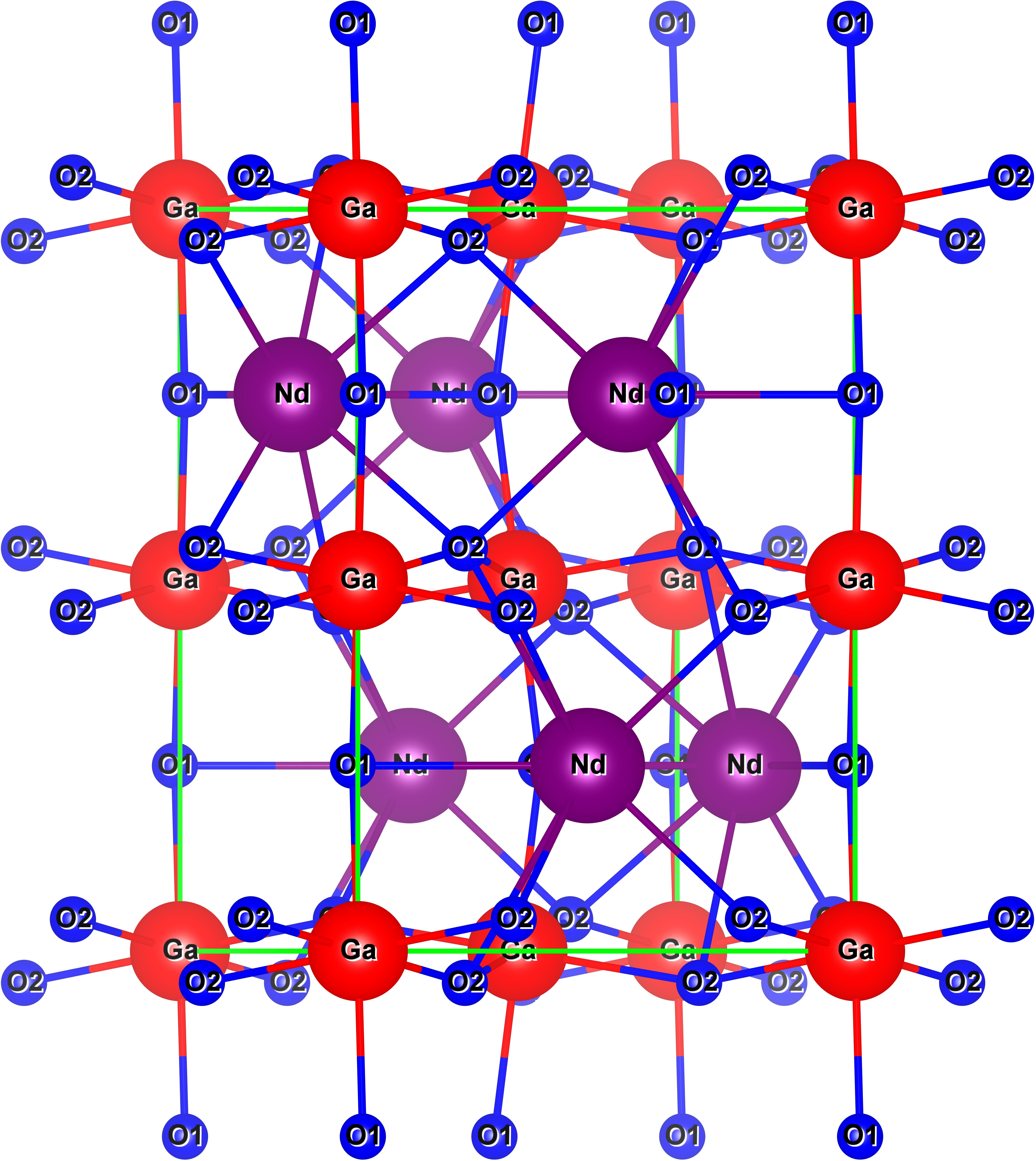}
\caption{\label{ngocrystal} (Color online) Orthorhombic crystal structure of NdGaO$_{3}$ (NGO) substrate.}
\end{figure}

The substrate NGO also has orthorhombic crystal structure with lattice parameters very close to the manganite film as shown in Fig.~\ref{ngocrystal}. The lattice parameters are given as\cite{Schmidbauer}

\begin{center}
$a$=5.428~A$^{\circ }$; \;\;\;\;\;\;\;\;\; $b$=5.498~A$^{\circ }$; \;\;\;\;\;\;\;\;\; $c$=7.709~A$^{\circ }$; \;\;\;\;\;\;\;\;\; $\alpha$ =$\beta$ =$\gamma$ =90$^{\circ }$
\end{center}

Since the lattice constants are well matched, there is negligible lattice strain (less than 0.1\%), so makes it ideal to grow high quality epitaxial homogeneous films. It further allowed us to safely ignore any strain related issues in the analysis of the measured data.\cite{Wuogale} 

\subsection{Temperature dependent optical and resistivity measurement}

Temperature dependent (20 K--300 K) reflectance measurements were conducted on both the film and substrate over a frequency range of 30-20000~cm$^{-1}$. Measurements in far infrared and mid infrared region were conducted using a Bruker 113v Fourier Transform Infrared spectrometer. A helium-cooled silicon bolometer detector was used in the 40-650~cm$^{-1}$ spectral range and a DTGS detector was used from 600-7000~cm$^{-1}$. Measurements in the near infrared and visible range were performed using a modified Perkin Elmer 16U grating spectrometer in conjunction with a continuous He flow cryostat. Room temperature reflectance data is extended up to 40,000~cm$^{-1}$ for the better understanding of high frequency behavior for both the substrate and the manganite film. All optical measurements were performed using non-polarized light at near-normal incidence geometry on the epitaxial grown film with exposed (010) plane. In order to complement the optical data, resistance measurement were also performed using standard four point probe technique both for cooling and heating cycles.  

\section{EXPERIMENTAL MEASUREMENTS AND ANALYSIS}
\subsection{Vibrational band structure and group theory analysis}
\begin{figure}[H]
\centering
\includegraphics[width=3.4 in,height=3.5 in,keepaspectratio]{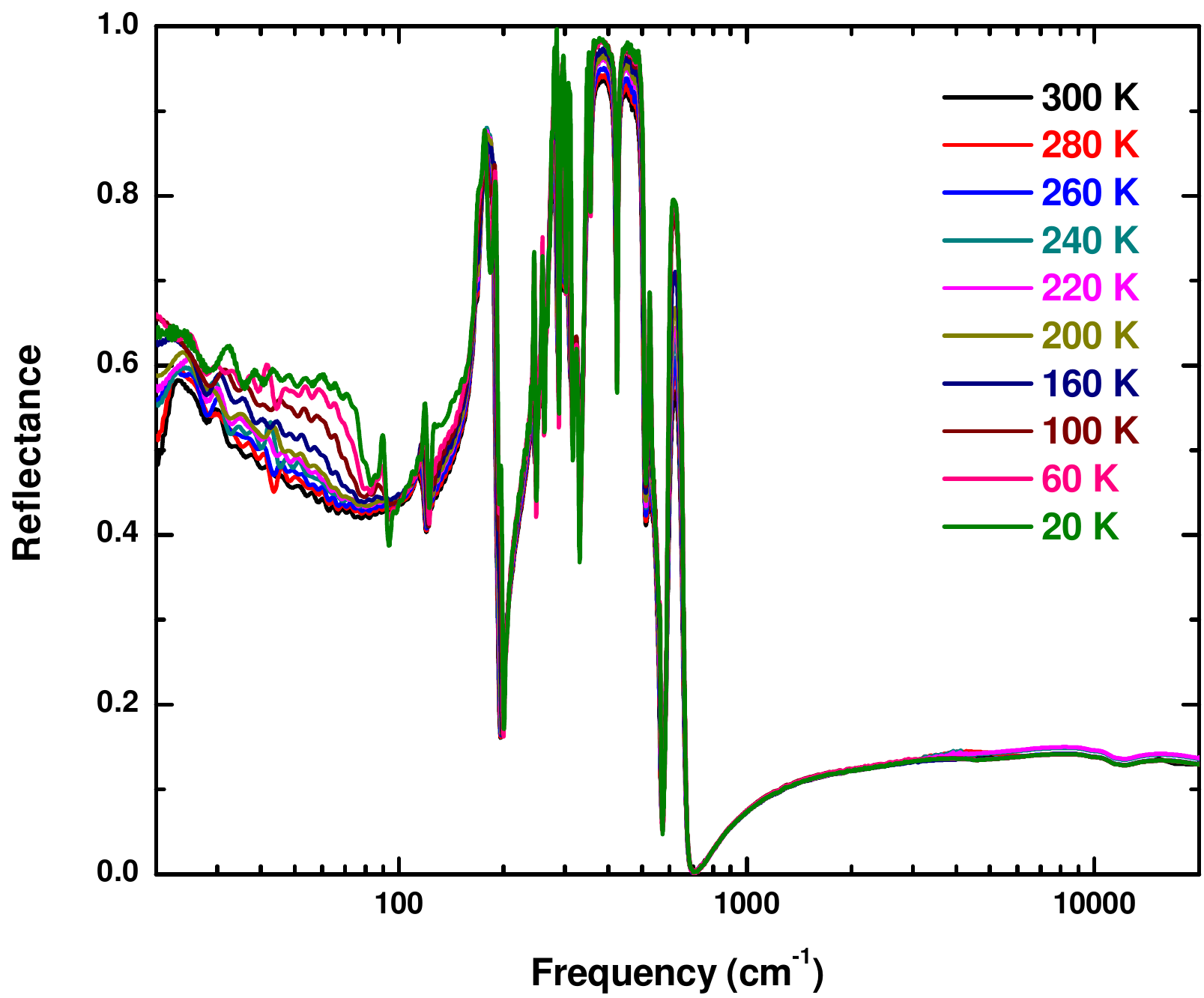}
\caption{\label{Reflectancesubstrate} (Color online) Temperature dependent reflectance spectra of NdGaO$_{3}$ (NGO) substrate.}
\end{figure}

In order to extract optical properties of La$_{0.67}$Ca$_{0.33}$MnO$_{3}$ film using exact formula for the thin film optics, it is necessary to quantify the substrate reflectance contribution first. The temperature dependent reflectance spectra of the NGO substrate at similar temperature points is shown in Fig.~\ref{Reflectancesubstrate}. We notice that the substrate reflectance does not show strong temperature dependence. 

The far-IR region is dominated by infrared active phonons at all temperatures. These phonon modes has also been reported and studied previously as a substrate for high-T$_{c}$ superconductive films.\cite{Calvani,Zhangzm} NGO has P$_{nma}$ (No. 62) space group symmetry. Based on the lattice parameters, atomic Wyckoff positions and lattice symmetry, the entire set of degrees of freedom is expressed in terms of the following irreducible point group representation.

\begin{equation}
\begin{aligned}
\Gamma_{3N}= {} & \textcolor{red}{7A_{g}+5B_{1g}+5B_{3g}+7B_{2g}}\\
& \textcolor{blue}{+ 9B_{1u}+9B_{3u}+7B_{2u}}\\
& \textcolor{green}{+ 8A_{u}+[1B_{1u}+1B_{3u}+1B_{2u}]}
\end{aligned}
\end{equation}

Here \textit{N} denotes total number of atoms in the periodic unit cell which is 20 (4 Nd, 4 Ga and 12 O). All modes in red color represent raman active modes (total 24 modes) while all in blue represent infrared active modes (total 25 modes). Rest 11 modes in green are optically inactive including 3 acoustic modes shown within the bracket. With an unpolarized light and near normal incidence along the epitaxial direction ($<$010$>$ or \textit{b}-axis), it should be possible to excite all optical phonons in the transverse \textit{ac}-plane. In our reflectance measurement, we are able to observe 17 optical phonons with B$_{1u}$ and B$_{3u}$ symmetry. Table~\ref{substrate parameters} lists all the experimentally observed vibrational parameters along with single electronic transition mode as derived from the Drude-Lorentz model fitting to the substrate reflectance data at 20 K as explained in the next section. It also shows 300 K parameters inside bracket for comparison.  

\begin{table}[h]
\caption {\label{substrate parameters}Drude-Lorentz  parameters for NGO substrate at 20 K(300 K parameters inside bracket).}
\centering
\begin{tabular}{cccc}
\hline\hline
  index  \;\;\;\;\; & $\omega _{pi}$  \;\;\;\;\; & $\omega _{0i}$ \;\;\;\;\;    & $\gamma _{i}$ \\
  \textit{i}      \;\;\;\;\;  & (cm$^{-1}$)    \;\;\;\;\;  &  (cm$^{-1}$)   \;\;\;\;\;    & (cm$^{-1}$)   \\[1ex]
\hline
  1      \;\;\;\;\;  &317(146)  \;\;\;\;\; &61(30) \;\;\;\;\; &47(36)\\[1ex]
\hline
  2      \;\;\;\;\;    &46(171) \;\;\;\;\; &90(70) \;\;\;\;\;  &2(66)\\[1ex]
\hline
  3      \;\;\;\;\;   &61(83)  \;\;\;\;\; &119(116) \;\;\;\;\;  &2(7)\\[1ex]
\hline
  4       \;\;\;\;\;   &489(393) \;\;\;\;\; &170(173) \;\;\;\;\;  &9(5)\\[1ex]
\hline
  5       \;\;\;\;\;   &110(87) \;\;\;\;\; &185(175)  \;\;\;\;\; &4(3)\\[1ex]
\hline
  6      \;\;\;\;\;  &227(114) \;\;\;\;\; &245(243) \;\;\;\;\; &2(3)\\[1ex]
\hline
  7       \;\;\;\;\;  &236(97) \;\;\;\;\; &259(256)\;\;\;\;\;  &2(2)\\[1ex]
	\hline
  8       \;\;\;\;\;  &643(562) \;\;\;\;\; &275(275) \;\;\;\;\; &2(6)\\[1ex]
	\hline
  9       \;\;\;\;\; &326(322) \;\;\;\;\; &291(290) \;\;\;\;\; &2(6)\\[1ex]
	\hline
  10      \;\;\;\;\;  &178(257)  \;\;\;\;\;&304(300) \;\;\;\;\; &3(10)\\[1ex]
	\hline
  11      \;\;\;\;\;  &247(388) \;\;\;\;\;  &321(319) \;\;\;\;\; &9(21)\\[1ex]
	\hline
  12     \;\;\;\;\; &528(546) \;\;\;\;\;&343(345) \;\;\;\;\; &4(11)\\[1ex]
	\hline
  13      \;\;\;\;\; &185(266) \;\;\;\;\; &356(356) \;\;\;\;\; &2(7)\\[1ex]
	\hline
  14     \;\;\;\;\;   &136(151) \;\;\;\;\; &425(424) \;\;\;\;\; &5(9)\\[1ex]
	\hline
  15     \;\;\;\;\;  &97(107) \;\;\;\;\; &516(516) \;\;\;\;\; &11(16)\\[1ex]
	\hline
  16      \;\;\;\;\;  &131(113) \;\;\;\;\; &547(539) \;\;\;\;\; &33(30)\\[1ex]
	\hline
  17       \;\;\;\;\; &227(259) \;\;\;\;\; &591(592) \;\;\;\;\; &11(27)\\[1ex]
	\hline
  18        \;\;\;\;\; &2766(2425) \;\;\;\;\;  &10695(10318) \;\;\;\;\; &3140(3155)\\[1ex]
\hline\hline
\end{tabular}
\end{table}

We notice that reflectance goes as high as 98\% between 275--525 cm$^{-1}$which then eventually decreases to almost zero just around the highest energy longitudinal mode appearing around 690 cm$^{-1}$. This far infrared feature generally denotes the presence of a restrahlen band which is quite commonly seen in polar crystals such as class III-V semiconductors InAs, GaN and GaAs\cite{Dmitruk,Jianyong,Fray}. This feature becomes even more pronounced as the temperature is lowered which generally indicates the suppression of phonon damping rate. The substrate becomes transparent above the plasma minimum, at around 700 cm$^{-1}$ beyond which substrate reflectance shows negligible frequency dependence, except around 11,500 cm$^{-1}$ which indicates f-f electronic transition (between $^{4}$I$_{j}$ ground states and $^{4}$F$_{j}$ excited j-multiplets) of Nd$^{3+}$ ion as previously reported.\cite{Martinhagan,Orera}.

\begin{figure}[H]
\centering
\includegraphics[width=3.4 in,height=3.5 in,keepaspectratio]{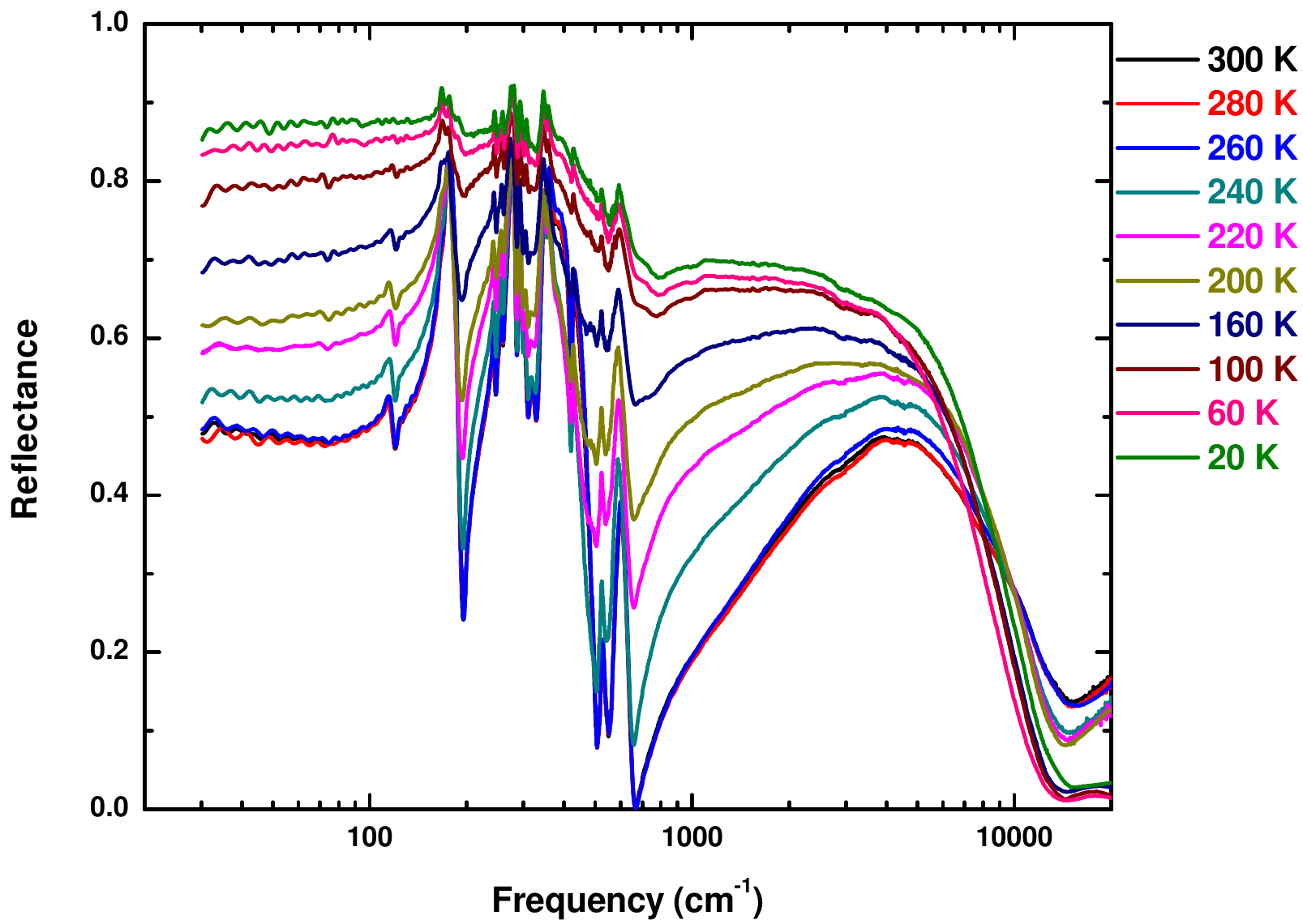}
\caption{\label{Reflectancelcmo} (Color online) Temperature dependent reflectance spectra of LCMO thin film on NGO substrate.}
\end{figure}

Now the temperature dependent reflectance spectra of La$_{0.67}$Ca$_{0.33}$MnO$_{3}$ film on NGO substrate is shown in Fig.~\ref{Reflectancelcmo} between 30 and 20,000 cm${}^{-1}$(4~meV--2.5~eV). The reflectance data of LCMO film on NGO substrate shows strong temperature dependence in all frequency regions except there is almost no change in measured reflectance between 260 K--300 K. As already mentioned, LCMO undergoes through a structural phase transformation from a high temperature distorted perovskite (orthorhombic) to a low temperature ideal perovskite (cubic) phase around that temperature range making its low temperature vibrational spectrum of prime interest. Based on the lattice parameters, atomic Wyckoff positions and lattice symmetry, the following irreducible point group representation gives the entire set of degrees of freedom for LCMO at low temperature.

\begin{equation}
\Gamma_{3N}=\textcolor{blue}{12T_{1u}}\textcolor{green}{+3T_{2u}+[3T_{1u}]}
\end{equation}

Here \textit{N}=6 denotes total number of atoms in the periodic unit cell (1 La with 0.67 occupancy, 1 Ca with 0.33 occupancy, 1 Mn and 3 O). All 12 modes in blue represent infrared active modes and rest 6 modes in green are optically inactive including 3 acoustic modes shown within the bracket. Owing to the cubic symmetry, all modes possess 3-fold degeneracy and hence we expect to see 4 distinct infrared phonon modes. Infrared region is dominated by phonons and strong phonon modes could be seen around 165 cm$^{-1}$, 340 cm$^{-1}$ and 562 cm$^{-1}$, apart from a weak mode around 520 cm$^{-1}$. Table~\ref{film parameters} lists all parameters for these phonons as derived from Drude-Lorentz model fitting.

\begin{table}[h]
\caption {\label{film parameters}Drude-Lorentz  parameters for LCMO film at 20 K.(300 K parameters inside bracket)}
\centering
\begin{tabular}{ccccc}
\hline\hline
  index  \;\;\;\;\; & $\omega _{pi}$  \;\;\;\;\; & $\omega _{0i}$ \;\;\;\;\;    & $\gamma _{i}$ \\
  \textit{i}      \;\;\;\;\;  & (cm$^{-1}$)    \;\;\;\;\;  &  (cm$^{-1}$)   \;\;\;\;\;    & (cm$^{-1}$)   \\[1ex]
\hline
  1      \;\;\;\;\;  &7240(275)  \;\;\;\;\; &0(0) \;\;\;\;\; &170(315)\\[1ex]
\hline
  2      \;\;\;\;\;  &2660(820)  \;\;\;\;\; &165(170) \;\;\;\;\; &74(77)\\[1ex]
\hline
  3      \;\;\;\;\;    &1160(2890) \;\;\;\;\; &340(330) \;\;\;\;\;  &20(130)\\[1ex]
\hline
  4      \;\;\;\;\;   &450(250)  \;\;\;\;\; &520(520) \;\;\;\;\;  &10(20)\\[1ex]
\hline
  5       \;\;\;\;\;   &630(1060) \;\;\;\;\; &562(575) \;\;\;\;\;  &24(85)\\[1ex]
\hline\hline
\end{tabular}
\end{table}

In addition, we noticed strong substrate reflectance features around 100 cm$^{-1}$ and 700 cm$^{-1}$. The visible region reflectance indicates a broad electronic transition mode or overlapping modes around 5000--8000 cm$^{-1}$ which is discussed later. In infrared region, reflectance increases as temperature is decreased while in the visible region, reflectance increases as temperature is increased. This feature suggests an electronic structure reconfiguration due to phase transition which is also discussed in later section.

\begin{figure}[H]
\centering
\includegraphics[width=3.5 in,height=3.5 in,keepaspectratio]{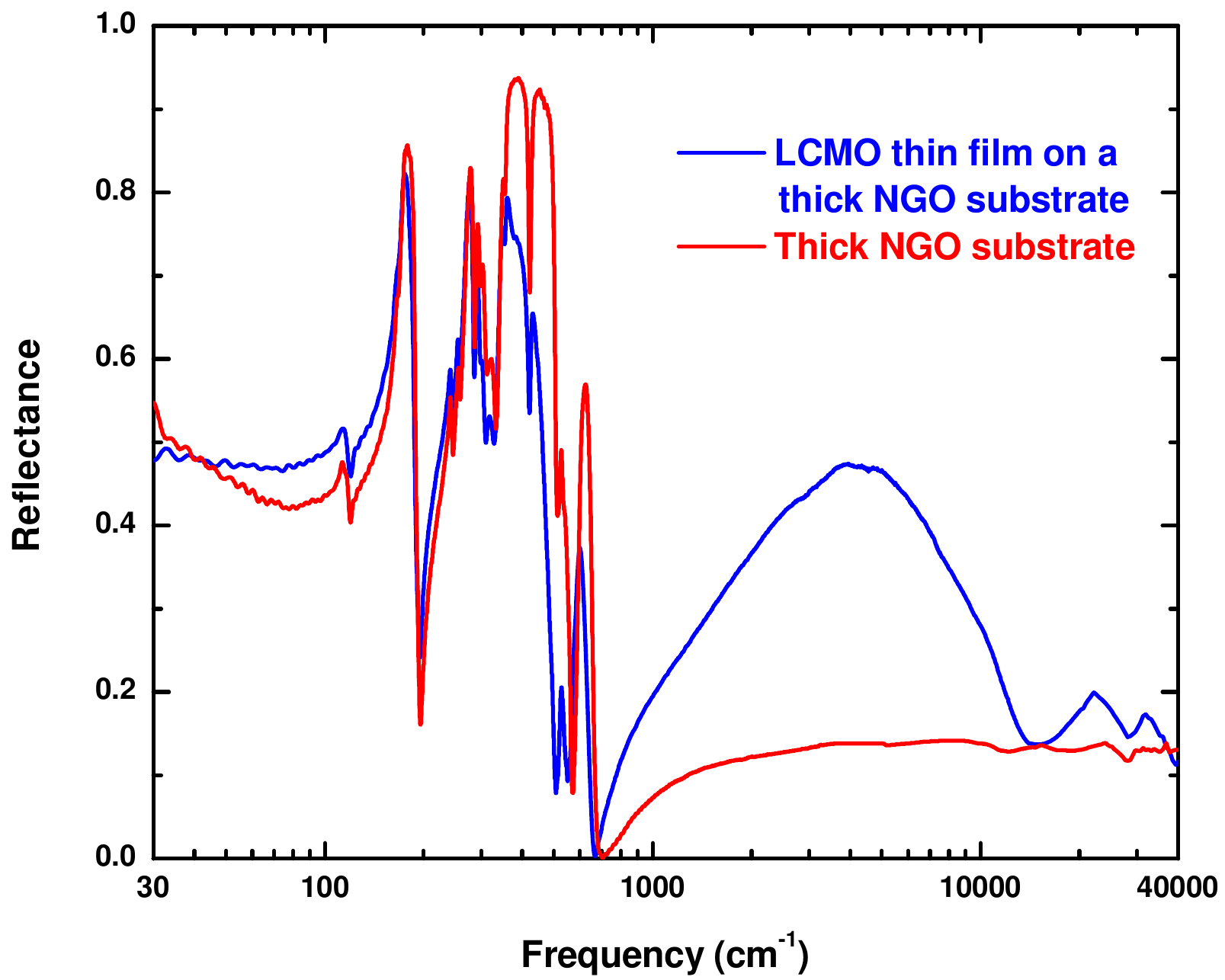}
\caption{\label{full300k} (Color online) Room temperature reflectance spectra of NGO substrate and LCMO film on substrate.}
\end{figure}
Room temperature reflectance spectra up to 40,000 cm$^{-1}$ is shown in Fig.~\ref{full300k} for both the substrate and film on substrate using photometric microscope. High frequency spectra shows more interband transitions for both the substrate and the film as expected.

\subsection{Drude-Lorentz and thin film optical analysis}

We performed Drude-Lorentz analysis on both the substrate and the film on substrate reflectance in order to derive other optical properties. The DL dielectric function is given as:

\begin{equation}
\varepsilon (\omega)=\varepsilon_{\infty}- \frac{\omega _{p}^{2} }{\omega^{2}+i\omega /\tau } +
\sum _{j=1}^N\frac{\omega_{pj}^{2} }{\omega_{j}^{2}-\omega^{2} -i\omega \gamma_{j} }
\end{equation}
where the first term represents the core electron contribution (transitions above the measured range, the second term is free carrier contribution characterized by Drude plasma frequency $\omega _{p}$ and free carrier relaxation time $\tau $ and the third term is the sum of several Lorentzian oscillators representing phonons, and interband electronic contributions. The Lorentzian parameters are the $j$th oscillator plasma frequency $\omega _{pj}$, its central frequency $\omega _{j}$, and its linewidth $\gamma _{j}$. This dielectric function model is used in a least-squares routine which minimizes the $\lambda^{2}$ between the acquired data and the computed one.\cite{Bevington}
After performing the Drude-Lorentz analysis on the substrate reflectance data as listed in Table~\ref{substrate parameters}, transfer matrix formalism is used to extract the DL fitting parameters for the LCMO layer which is listed in Table~\ref{film parameters}. The transfer matrix for the k$^{th}$ layer of thickness t$_{k}$ and complex dielectric function $\varepsilon_{k}$ at incident wavelength $\lambda$ is given as

\begin{subequations}
\begin{align}
{} & M_{k}=\begin{pmatrix}
\cos\delta_{k}  &-\frac{i}{\sqrt{\varepsilon_{k} }}\sin\delta_{k} \\ 
-{i}{\sqrt{\varepsilon_{k} }}\sin\delta_{k} & \cos\delta_{k}
\end{pmatrix}\\
& \delta_{k}=(2\pi/\lambda )\sqrt{\varepsilon _{k}}{t_{k}}
\label{Eq5}
\end{align}
\end{subequations}
The overall reflection from a multilayer system depends on the product of all such transfer matrices taken in sequence. If the product of film transfer matrix M$_{f}$ (unknown) and the substrate transfer matrix M$_{s}$ (known) is represented as shown below in Eq.~\ref{Eq6a}, then the complex reflection coefficient and the power reflectance from the film over substrate could be written subsequently in terms of unknown complex transfer matrix elements as shown in Eq.~\ref{Eq6c}.

\begin{subequations}
\begin{align}
{} & M_{f}M_{s}=\begin{pmatrix}
A  & B \\ 
C &  D
\end{pmatrix}\label{Eq6a}\\
& r_{total}=\frac{A+B-C-D}{A+B+C+D}\\
& R_{total}=r_{total}(r_{total})^{*}= \begin{vmatrix}
\frac{A+B-C-D}{A+B+C+D}
\end{vmatrix}^{2}
\label{Eq6c}
\end{align}
\end{subequations}

\begin{figure}[H]
\centering
\includegraphics[width=3.2 in,height=3.2 in,keepaspectratio]{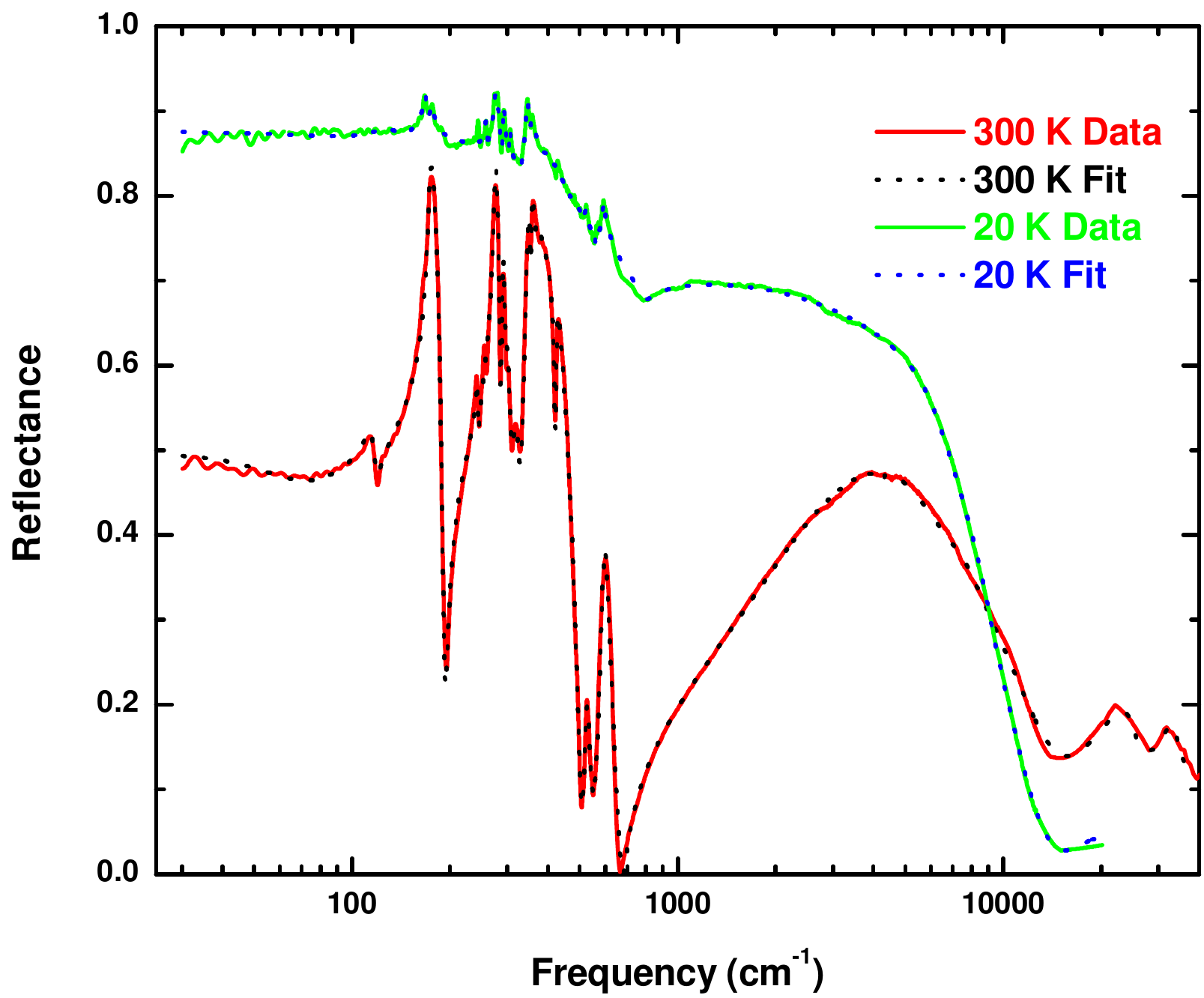}
\caption{\label{fitlcmo} (Color online) Drude-Lorentz fit at 20 K and 300 K reflectance spectra of LCMO film on NGO.}
\end{figure}

Figure~\ref{fitlcmo} shows the comparison between measured reflectance and DL calculated reflectance spectra of LCMO film on substrate at base temperature of 20 K and at room temperature of 300 K. Similar quality fitting procedure is repeated at all other intermediate temperatures in order to analyze the temperature dependence of several optical properties of LCMO.
  
\subsection{Optical and transport properties}

Real part of the optical conductivity $\sigma_1(\omega)$ and their temperature dependence over the measured frequency range has been calculated from the Drude-Lorentz fitting parameters using Eq.~\ref{Eq7}

\begin{equation}
\sigma_1 (\omega)=\frac{\omega _{p}^{2}\tau }{4\pi[1+\omega^2 \tau^2]} +
\sum _{j=1}^N\frac{\omega_{pj}^{2}\omega^{2}\gamma_{j} }{4\pi[(\omega_{j}^{2}-\omega^{2})^2 +\omega^2 \gamma_{j}^2] }
\label{Eq7}
\end{equation}

\begin{figure}[H]
\centering
\includegraphics[width=3.3 in,height=3.3 in,keepaspectratio]{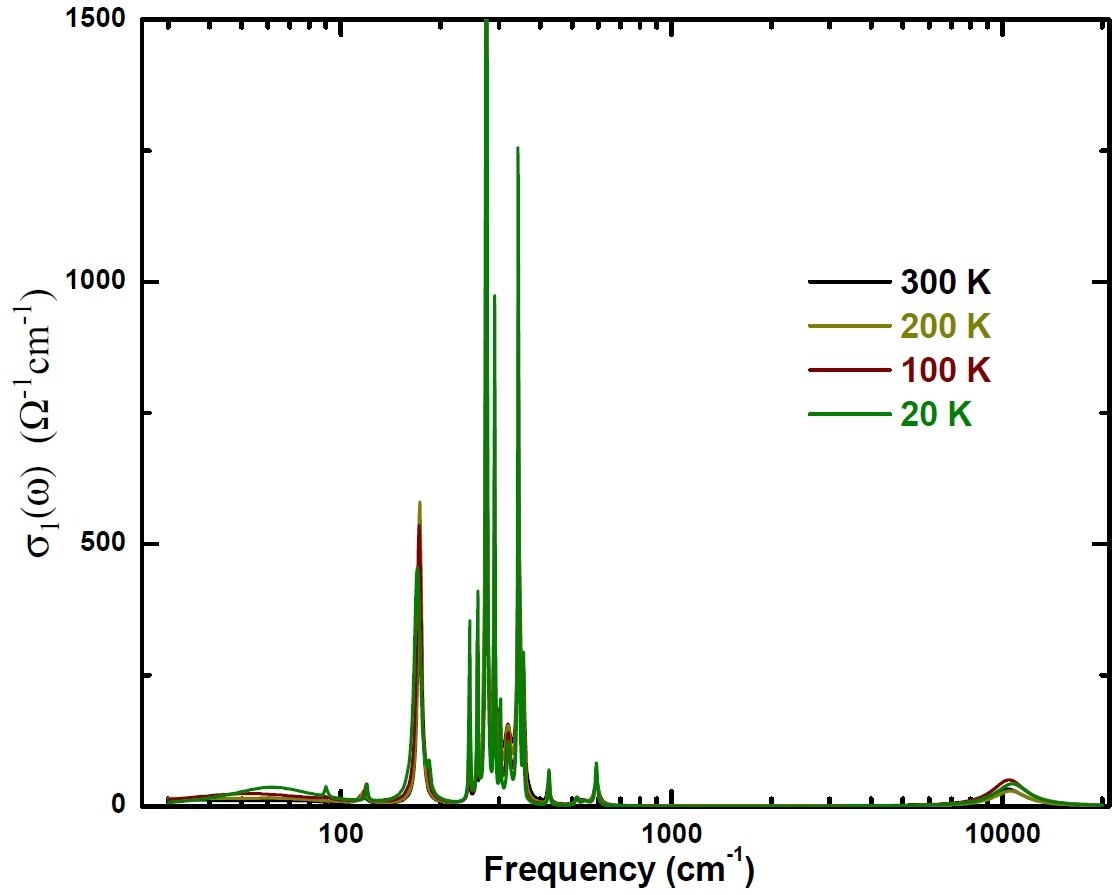}
\caption{\label{conductivityngo} (Color online) Temperature dependence of DL parameter calculated $\sigma_1(\omega)$ of NGO.}
\end{figure}

As we see in Fig.~\ref{conductivityngo} that substrate remains an insulator and shows all far-infrared phonon modes as previously listed in Table~\ref{substrate parameters}. Higher frequency spectra show electronic transition however no temperature dependence is observed in any measured frequency regions as expected from a weakly temperature dependent reflectance spectra. 

\begin{figure}[H]
\centering
\includegraphics[width=3.4 in,height=3.4 in,keepaspectratio]{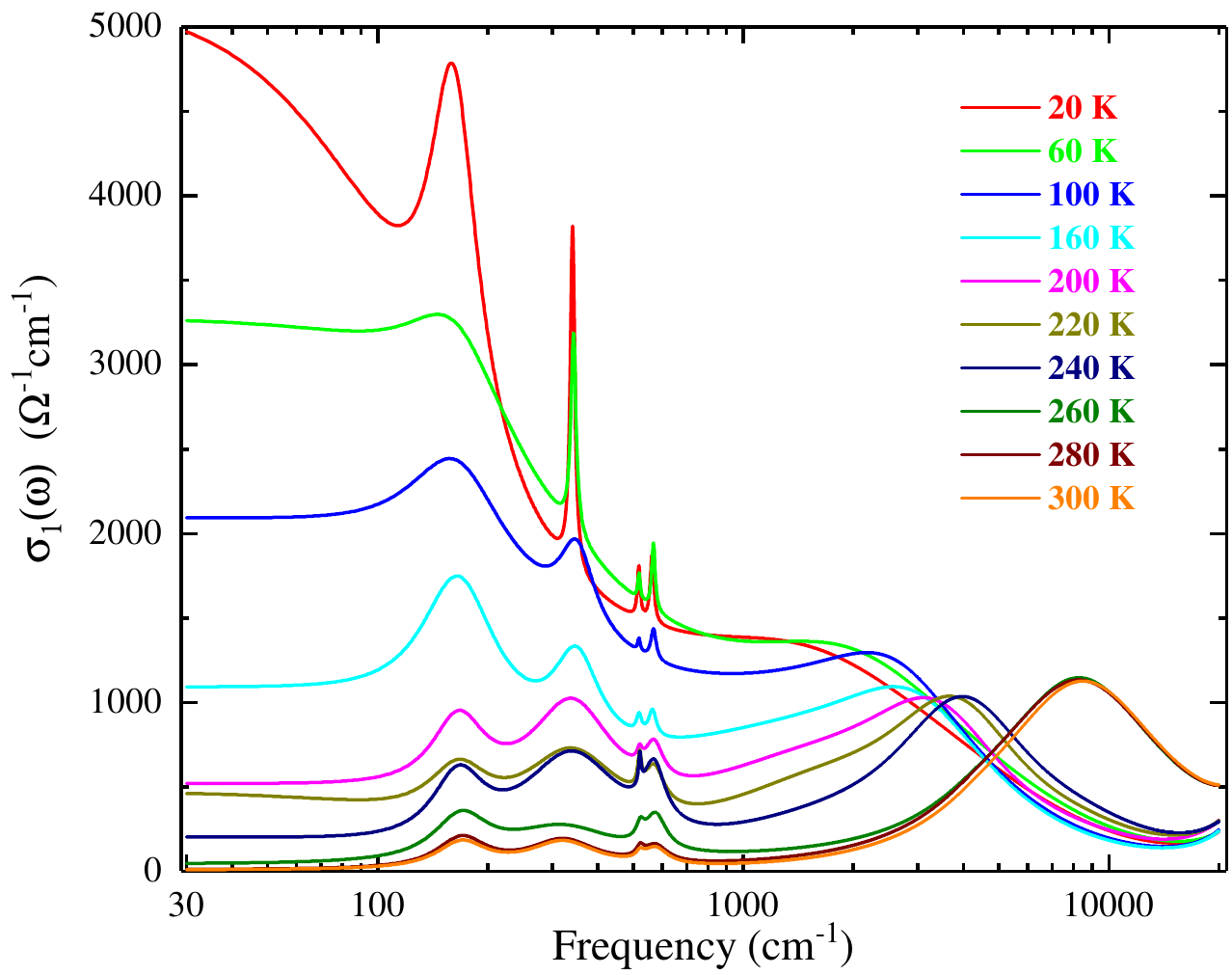}
\caption{\label{conductivitylcmo} (Color online) Temperature dependence of DL parameter calculated $\sigma_1(\omega)$ of LCMO film.}
\end{figure}

On the other hand, LCMO film conductivity plot in Fig.~\ref{conductivitylcmo} shows strong temperature dependence in all frequency regions. The far infrared region is dominated by optically active phonons superimposed over Drude conductivity contribution both showing strong temperaure dependence. Far infrared conductivity keeps on decreasing monotonically as temperature increases upto 260 K and above which DC conductivity seems to be independent of temperature.

\subsubsection{Characterization of charge carrier dynamics}
\begin{figure}[H]
\centering
\includegraphics[width=3.2 in,height=3.2 in,keepaspectratio]{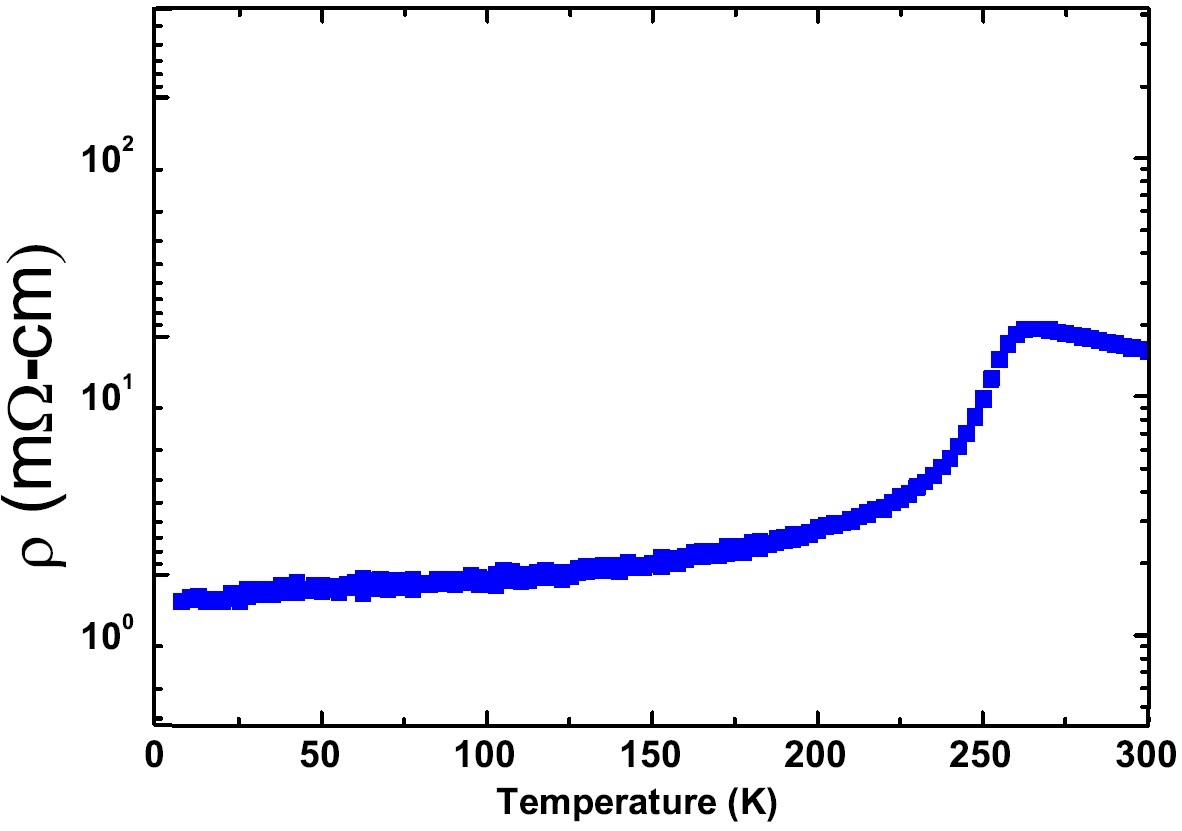}
\caption{\label{LCMODCresistivity} (Color online) Temperature dependence of LCMO film resistivity using transport technique.}
\end{figure}

Film DC resistivity as a function of temperature is measured using standard four-point probe transport technique during the cooling and the heating cycle reflects similar trend as shown in Fig.~\ref{LCMODCresistivity}. The resistivity starts dropping around 260 K during the cooling cycle and keeps decreasing at noticable rate upto 230 K. This behavior is repeated during the heating cycle and no hysteric behavior is noticed. This resistivity peak indicates the presence of a first order phase transition from insulator to metal phase in the LCMO film. Further qualitative comparison between the transport dc conductivity and the optical reflectance is shown in the far infrared

\begin{figure}[H]
\centering
\includegraphics[width=3.35 in,height=4.0 in]{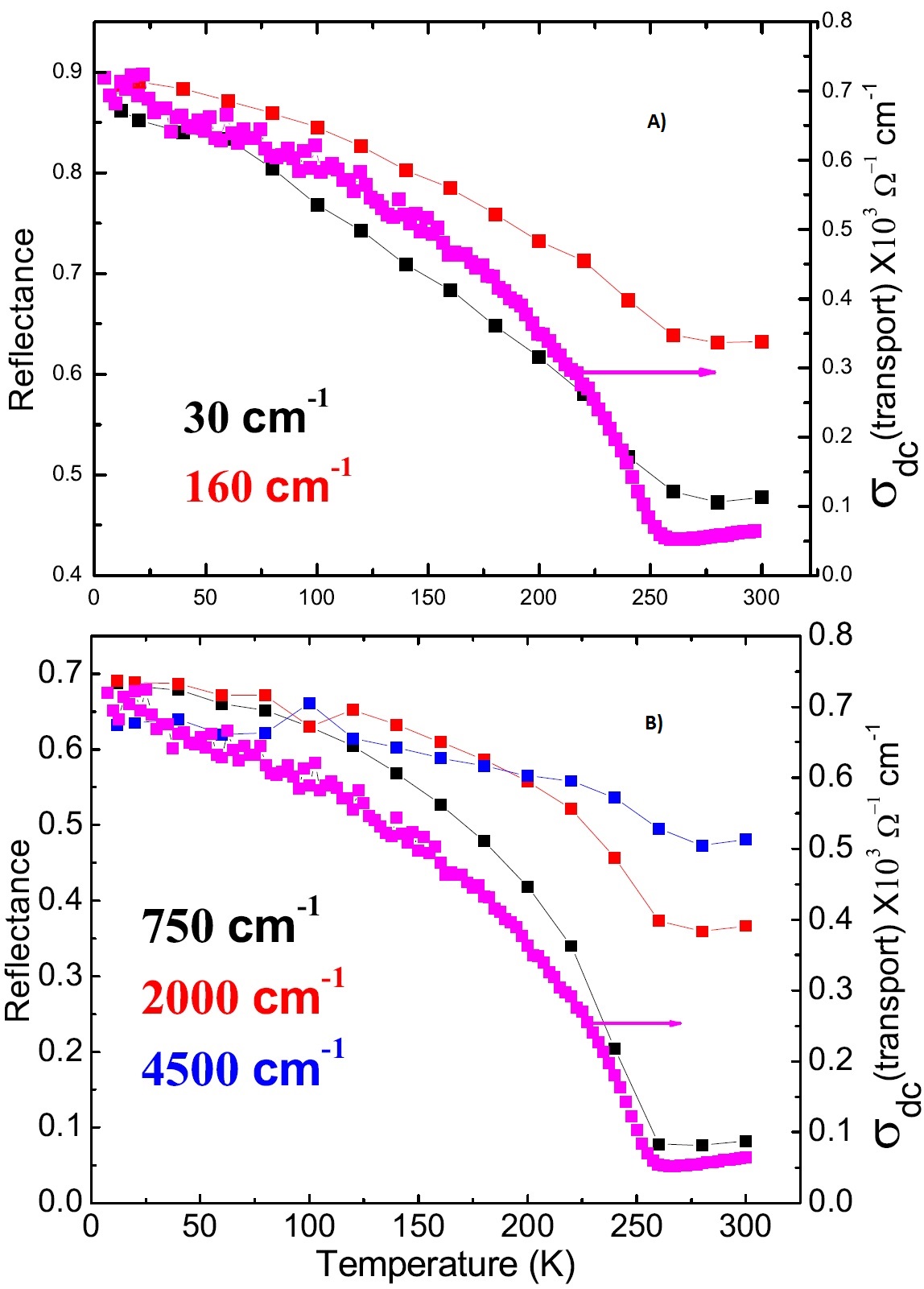}
\caption{\label{opticstransport} (Color online) Comparison between the measured optical reflectance and transport dc conductivity in A) far-infrared B) mid-infrared range.}
\end{figure}

\begin{figure}[H]
\centering
\includegraphics[width=3.4 in,height=3.5 in]{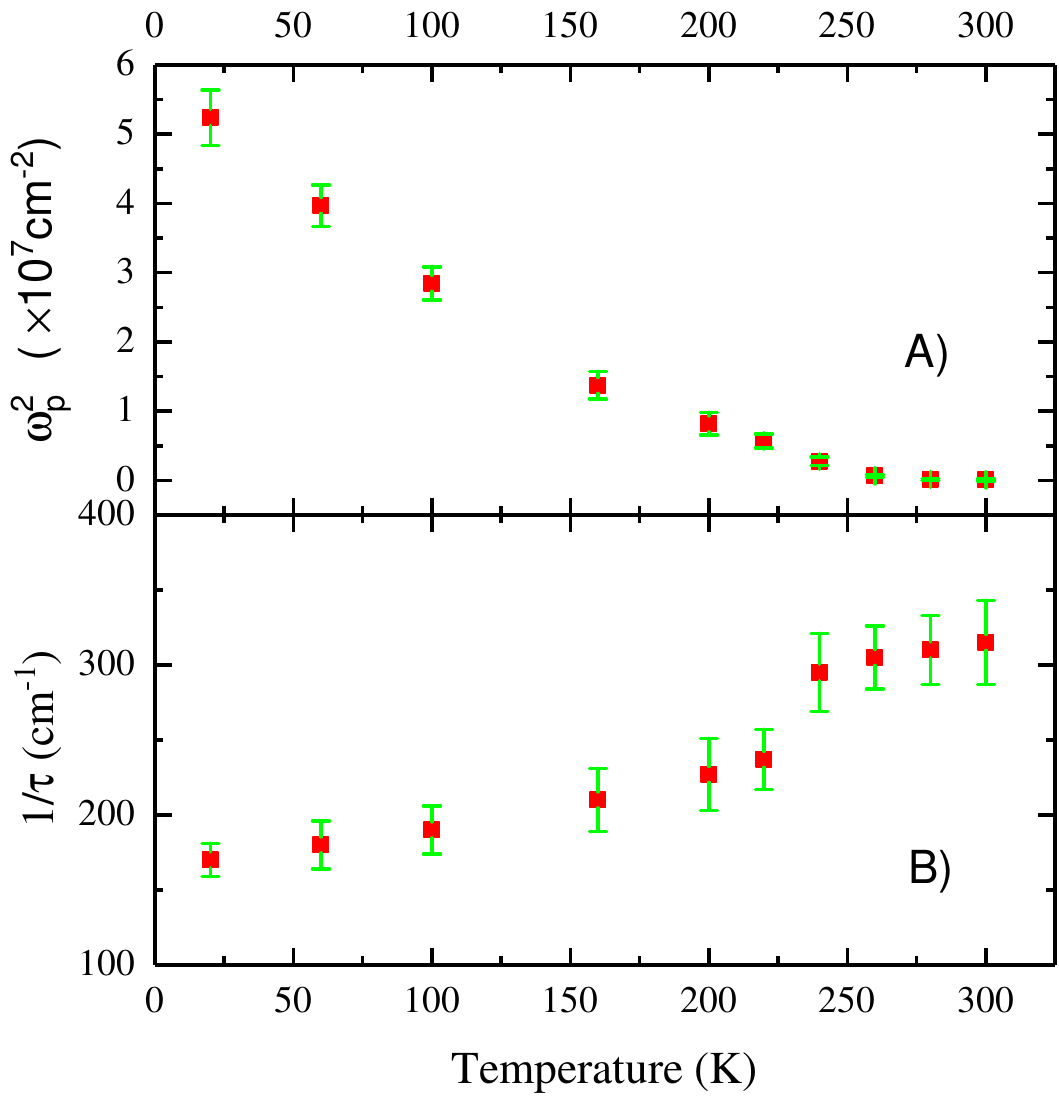}
\caption{\label{carrier} (Color online) Temperature dependence of A) Drude strength and B) scattering rate of LCMO film.}
\end{figure}

\hspace{-0.4 cm}(panel A) and mid infrared range (panel B) of Fig.~\ref{opticstransport}. In both panels one can see a concomitant increase in the measured reflectance and dc conductivity as temperature goes below the metal-insulator transition temperature of 260 K. The change in the reflectance and the sharpness as measured at 4500 cm$^{-1}$ is not as big in comparison to other lower frequencies but overall trend is qualitatively the same. In addition, negligible change in the refelctivity and dc conductivity is noticed above the T$_{MI}$ in both panels. To get further insight into the dynamics of charge carriers in LCMO film, the Drude plasma strength and charge carrier scattering rate has been plotted against temperature in Fig.~\ref{carrier}. The experimental values of $\omega_{p}^{2}$ is approximately proportional to (T$_{MI}$ -- T) in the metallic region which is an indication of a progressive increase in density of states at the Fermi energy upon cooling.\cite{ParkChen} The scattering rate however increases non linearly with temperature. It changes slowly above base temperature within the error bars but starts increasing more rapidly as temperature increase above 200 K and almost saturates in the insulator region of measurement. If we assume that carrier density \textit{n} is equal to 0.3 hole per Mn at 20 K, one can estimate the mean free path $\textit{l}=\frac{(3\pi^2)^{1/3}\hbar\sigma_{DC}}{e^{2}n^{2/3}}$$~\approx$~22~A$^{\circ }$. In comparison to the lattice constant of 3.9~A$^{\circ }$, mean free path is much larger which suggests that the origin of the metallic nature in LCMO film is due to the coherent motion of a large polaron appearing at frequencies below the characteristic phonon modes.\cite{Emin} 

A simple comparison between infrared and transport conductivity (Fig.~\ref{conductivitylcmo} and Fig.~\ref{opticstransport}) however shows some sizable difference. Such discrepency has been investigated for polycrystalline La$_{0.7}$Ca$_{0.3}$MnO$_{3}$ samples in the past.\cite{KimGu} The higher resistivity in the transport measurement is attributed to the influence of intergranular resistance which is predominant channel of carrier flow in the transport measurement. High and low resistive regions are connected in series and the dc response strongly depends on the high resistive region.\cite{KimGu} On the other hand, infrared response reflects more of the intrinsic properties of grains. It repsents an average over a length scale of $\lambda$/\textit{n} where $\lambda$ is the wavelength of the incident radiation and \textit{n} is the refractive index of the material. Following this line of reasoning, at 20 K, the refractive index of the film was found close to 50 around 50 cm$^{-1}$ which makes the infrared length scale around 4 micron. However, previous studies have reported larger grain size in LCMO films over NGO.\cite{AmlanBiswas} which makes the IR response insensitive to the intergrain resistance. Besides, IR resitivity shows stronger temperature dependence below the critical temperature as compared to the transport resistivity. If the difference in terms of the predominant free carrier scattering channel between optical and transport measurements were to be believed, it is obvious that IR response provides intrinsic electrodynamic properties within the grains and thefore a better microscopic representation of free carrier dynamics and its temperature dependence.

\subsubsection{Jahn-Teller distortion and first order phase transition}
 
\begin{figure}[H]
\centering
\includegraphics[width=3.3 in,height=3.6 in]{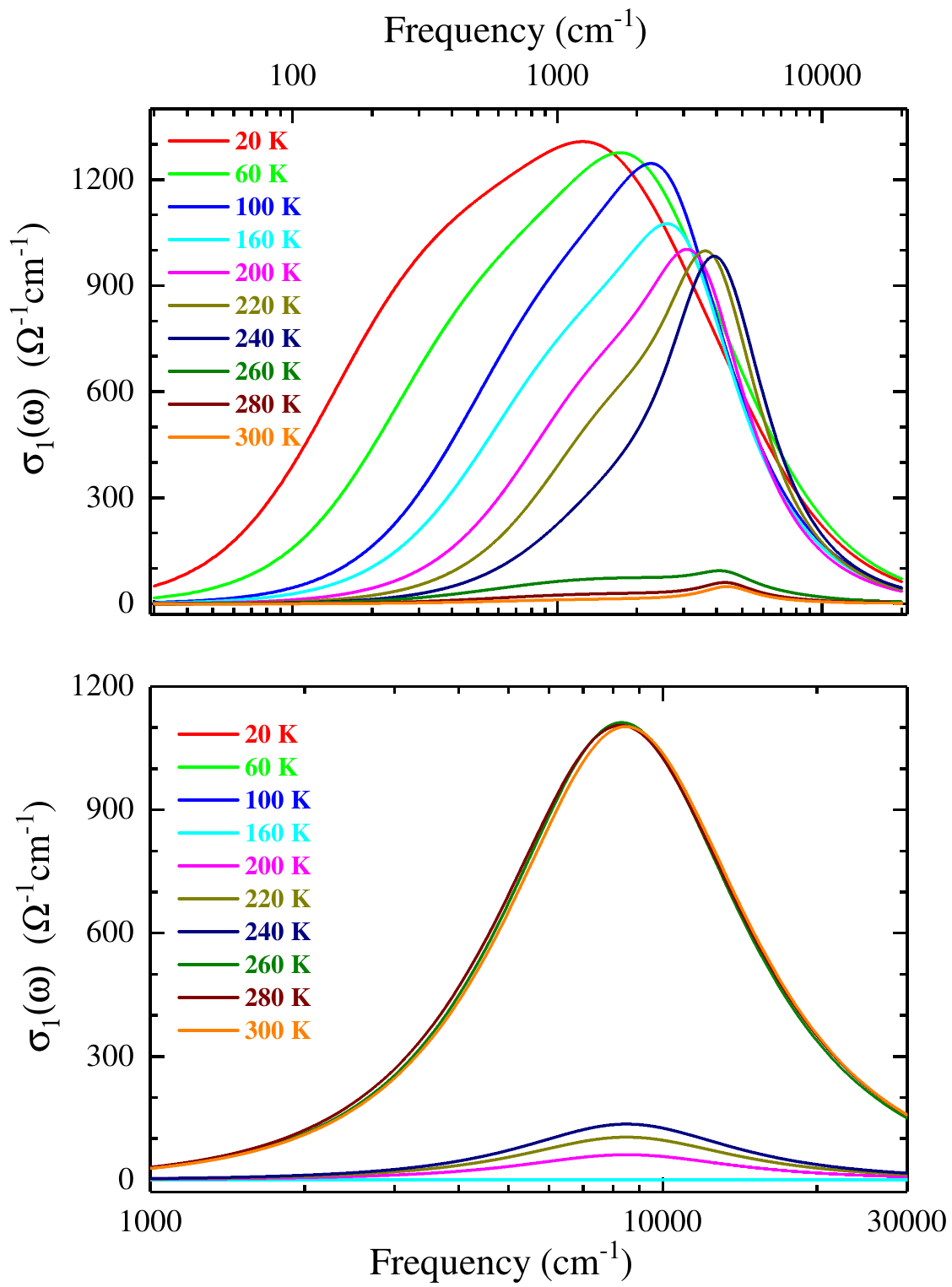}
\caption{\label{midIR} (Color online) Temperature dependence of A) incoherent mid-IR polaron bands in metallic phase and B) Holstein polaron in insulating phase of LCMO film.}
\end{figure}

In the mid-IR to visible range, several overlapping absorption bands are located contributing towards the optical conductivity. Fig.~\ref{midIR} panel A) shows the temperature evolution of two overlapping mid-IR absorption bands and their conductivity contribution in the metallic phase. 
Theoretical model of frequency response of large polaron photoionization predicts an incoherent and very asymmetric mid-IR band with a long tail above its peak position which is quite evident in the figure.\cite{Emin,Kimjung}  Besides having larger asymmetry at lower temperature, the peak position shifts towards higher frequency and the band strength decreases as temperature is increased. Above the critical temperature of 260 K, strength is almost negligible and all the spectral weight shifts to even higher frequency centered around 8000~cm$^{-1}$ and spread over much wider frequency range as shown in panel B) of Fig.~\ref{midIR}. Because of the octahedral symmetry around Mn sites, the configuration becomes $t_{2g}^{3}e_{g}^{1}(^{5}E)$ for the Mn$^{3+}$ and  $t_{2g}^{3}(^{4}A_{2})$ for the Mn$^{4+}$ sites. Moreover, due to the asymmetric occupation of degenerate $e_{g}$ orbitals at Mn$^{3+}$ sites, system tries to get rid of the extra energy by lowering the overall symmetry of the molecule by undergoing distortion known as Jahn Teller distortion. The Z-out JT distortion lowers the octahedral symmetry to tetragonal symmetry around the Mn sites which brings the energies of all orbitals with z-factor ($d_{xz}$, $d_{yz}$, $d_{z^{2}}$) lower creating a small indirect gap between $e_{g}$ states, with $d_{x^{2}-y^{2}}$ orbital character being unoccupied and higher in energy. Theoretical calculation has predicted large width of these splitted energy levels due to unstable and non-uniform JT distortion\cite{Satpathy,SatpathyPopov} explaining the wide nature of absorption bands in our experimental results. On the other hand, JT distortion is very small or negligible at Mn$^{4+}$ sites. 

In the low temperature metallic state, JT distortion is small and octahedral symmetry is maintained which preserves the degeneracy of two $e_{g}$ states. Moreover, mobile electrons in the $e_{g}^{1}$ bands couple the $t_{2g}^{3}$ spins (\textit{S}=3/2) ferromagnetically via double exchange which leads to coexisting long range magnetic ordering in the metallic phase. It has a significant imapct on lattice-electron interplay which extends the coupling between them beyond a single lattice site leading to the formation of large polarons. Below the characteristic phonon frequency, this polaron shows Drude like behavior as explained in previous section and characterized in Fig.~\ref{carrier}. Above the characteristic phonon frequency, polaron photoionizes and shows a asymmetric band with long tail as shown in panel A) of Fig.~\ref{midIR}. In the high temperature paramagnetic insulative phase, JT distortion is significant at Mn$^{3+}$ sites. Symmetry of the system is lowered to tetragonal or orthorhombic which leads to the rearrangement of energy levels. Breaking of degeneracy creats an indirect gap between $e_{g}$ states on Mn$^{3+}$ sites leading to the formation of small localized polarons known as Holstein polarons as also reported in other similar perovskites.\cite{Okimoto,Quijada,Kimjung} This electronic rearrangement explains the conductivity peak around 1 eV due to intra-atomic $e_{g}^{1}$(Mn$^{3+}$)~$\rightarrow$~ $e_{g}^{2}$(Mn$^{3+}$) transition and interatomic $e_{g}^{1}$(Mn$^{3+}$)~$\rightarrow$~ $e_{g}$(Mn$^{4+}$) transition\cite{Jungkim,Arima,Saitoh} in the visible range when the system is in insulative phase as shown in panel B) of Fig.~\ref{midIR}. Although electric dipole selection rule prohibits intra-atomic $\textit{d}~\rightarrow~\textit{d}$ type transition but strong hybridization of $e_{g}$ bands with O 2\textit{p} bands and noncubic distorted local symmetry of the MnO$_{6}$ octahedron due to JT distortion will allow such transition away from the $\Gamma (\textit{k}=0)$ point.\cite{Eom,Chainani}  

\subsubsection{Evolution of phonon structure under strong electron-lattice coupling}

\begin{figure}[H]
\centering
\includegraphics[width=3.5 in,height=4 in]{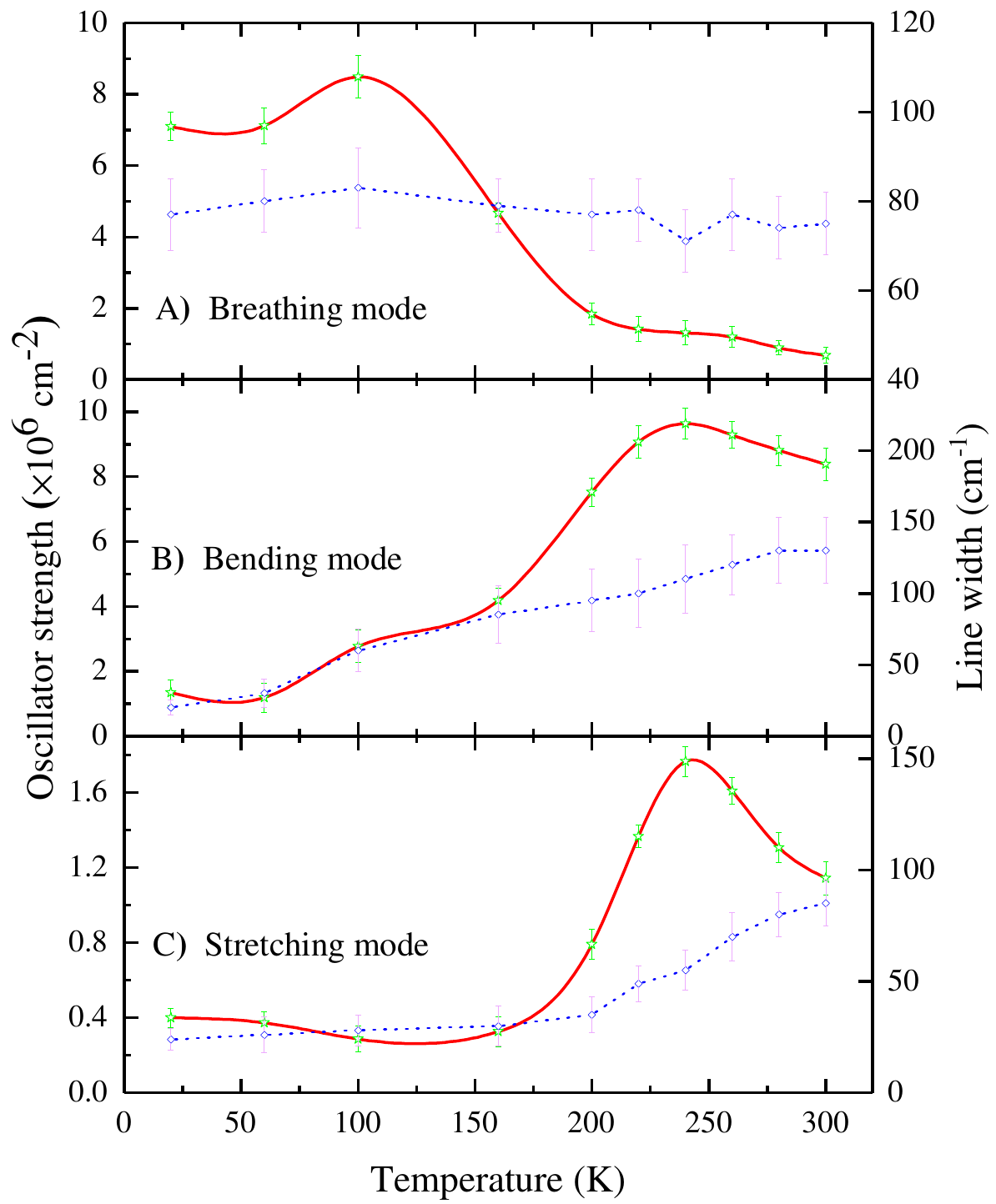}
\caption{\label{phonons} (Color online) Temperature dependence of oscillator strength (solid line) and line width (dot line) of Jahn-Teller modes in LCMO film.}
\end{figure}

On account of earlier cited crossover from a low temperature coherent transport in ferromagnetic phase to a high temperature Jahn-Teller distortion localized small polaronic paramagnetic phase, it is of great interest to follow the evolution of phonon structures across the phase transition. Due to the presence of strong electron-lattice coupling in  perovskites, polaronic transport under existing local distortions of the lattice is bound to influence the shape and the strength of associated vibrational modes. Far-IR region of the measured reflectance (Fig.~\ref{Reflectancelcmo}) and DL derived optical conductivity (Fig.~\ref{conductivitylcmo}) data show several temperature dependent phonon modes which is listed in Table~\ref{film parameters}.  The external mode or breathing mode around 165 cm$^{-1}$ represents a vibrating motion of Ca$^{2+}$ or La$^{3+}$ cations against MnO$_{6}$ octahedral unit. This mode couples to the changes in the $e_{g}$ charge density and determines the electron hopping amplitude for a given choice of cations and stoichiometry $x$. The bending mode or the basal-plane distortion mode at 340 cm$^{-1}$ reflects an internal motion of the manganese and oxygen ions located in the basal plane against the other oxygen ions in a plane perpendicular to it. On the other hand, the stretching mode at 562 cm$^{-1}$ corresponds to another internal motion of manganese ions against the oxygen octahedron leading to the compression in Mn-O bonds in the basal \textit{xy}-plane and elongation in Mn-O bonds along the \textit{z}-direction. These two internal modes of MnO$_{6}$ octahedron couple to preferential occupancy of of one type of $e_{g}$ orbital ($d_{z^{2}}$) over the other $e_{g}$ orbital ($d_{x^{2}-y^{2}}$) during the structural phase transition which eventually prompts the octahedron to develop two different O sites in the high temperature phase. The Jahn Teller distortion is essentially a combination of these vibrational modes and Fig.~\ref{phonons} shows the temperature dependence of the oscillator strengths of all three modes across the first order phase transition around 260~K. 

The breathing mode in panel A) around 165 cm$^{-1}$ shows higher strength in low temperature metallic phase (a weak minima around 100 K) and decreases by an order of magnitude in the high temperature insulative phase. This behavior suggests that large polaron metallic phase favors for stronger coupling between divalent or trivalent cations with the MnO$_{6}$ octahedral units and the coupling strength decreases as such polarons get localized in the high temperature insulative phase. On the contrary, the bending mode in panel B) around 340 cm$^{-1}$ shows higher oscillator strength in the high temperature insulative phase which drops by an order of magnitude as temperature decreases below the critical temperature of 260~K. Besides, this mode appears to have much larger linewidth in the insulative phase which is expected on account of local structural variations due to inhomogeneous distribution of Mn$^{3+}$ and Mn$^{4+}$ ions in the insulative phase. Nonetheless, the impact of two discrete Mn ion environments gets smeared and linewidth decreases due to the mobility of large polarons in the metallic phase. On the other hand, the stretching mode in panel C) around 562 cm$^{-1}$ shows a clear maxima around the critical temperature and then oscillator strength quickly decreases to about one quarter of the maximum value in the low temperature metallic phase. The mid-IR polaronic band in the low temperature metallic phase seems to weaken the coupling strength of the stretching mode through charge delocalization therefore, adversely influencing the tendency of preferential occupancy of $d_{z^{2}}$-type $e_{g}$ orbitals in the high temperature insulative phase. Besides, the asymmetric shape of this mode with a stretched tail at high frequency side (Fig.~\ref{conductivitylcmo}) points towards phonon-polaron interaction with coexisting asymmetric polaronic band at high frequency side in the metallic phase. Decreasing polaronic background at higher temperature supresses the electronic screening of the stretching mode and phonon strength increases.\cite{Premila} The weak mode around 520 cm$^{-1}$ does not show any temperature dependence.

\section{CONCLUSIONS}
Temperature dependent reflectance measurement on LCMO film and NGO substrate has allowed to study the evolution of key dynamical components which modulates the electronic and magnetic behaviour in the manganite system across the phase transition. The Drude-Lorentz model along with thin-film optical analysis has enabled us to examine the temperature evolution of Jahn-Teller modes in manganites in terms of their changing oscillator strengths and line widths. Low temperature metallic phase is dominated by heavy polaron dynamics and key dynamical variables such as scattering rate, Drude strength and mean free path has been estimated. The interaction of incoherent and asymmetric polaronic background in the mid-IR range with the Jahn-Teller modes has been qualitatively discussed. Localization of free carriers as Holstein polarons with the growing Jahn-Teller distortion in the high temperature insulative phase is explained in terms of the rearrangement in the electronic structure. It leads to the formation of an indirect bandgap which is further supported by the optical conductivity calculations with emerging conductivity peak in the visible range only in the high temperature insulative phase.
\bibliography{LCMO}
\end{document}